\documentstyle[aps,manuscript,epsf,eqsecnum]{revtex}  

\newcommand{\bxi}[0]{\mbox{\boldmath $\xi$}}
\newcommand{\grad}[0]{\mbox{\boldmath $\nabla$}}
\newcommand{\eps}[0]{\varepsilon}
\newcommand{\bx}[0]{{\bf x}}
\newcommand{\by}[0]{{\bf y}}
\newcommand{\br}[0]{{\bf r}}

\newcommand{\tg}[0]{\widetilde{g} }
\newcommand{\tc}[0]{\widetilde{c} }

\newcommand{\tJ}[0]{\widetilde{J} }
\newcommand{\hG}[0]{\widehat{G} }

\title{Disorder Induced Depinning Transition}
\author{Terence Hwa~\cite{addr}}
\address{School of Natural Sciences\\
Institute for Advanced Study\\
Princeton, NJ  08540, U.S.A.}
\author{Thomas Nattermann}
\address{Institut f\"ur Theoretische Physik\\
Universit\"at zu K\"oln\\
D-50937 K\"oln, Germany}
\date{\today}
\begin{document}
\maketitle
\begin{abstract}
The competition in the pinning of a directed polymer by a columnar pin
and a background of random point impurities is investigated systematically
using the renormalization group method. With the aid of the mapping to the 
noisy-Burgers' equation and the use of the mode-coupling method, the directed 
polymer is shown to be marginally localized to an arbitrary weak columnar pin 
in $1+1$ dimensions. This weak localization effect is attributed to the 
existence of large scale, nearly degenerate optimal paths of the randomly 
pinned directed polymer. The critical behavior of the depinning transition 
above $1+1$ dimensions is obtained via an $\epsilon$-expansion. 
\end{abstract}
\vfill
\narrowtext

\newpage
\section{Introduction}

The statistical mechanics of an elastic manifold embedded in  a medium 
of random point defects has been
the subject of many studies in the past decade~\cite{nat0,hh,kz}. 
Such systems are encountered in
a variety of contexts,  ranging from the fluctuations of domain walls in
random magnets~\cite{nat0,hh,dw,manifold}, 
to the pinning of magnetic flux lines in dirty
superconductors~\cite{creep,nat1,bmy}. Over the years, many theoretical 
methods have been developed to understand these
systems~\cite{cayley,migdal,bethe,p,mp,fh,hhf,iv,hf}. In particular,
through a mapping to stochastic hydrodynamics~\cite{hhf,iv,hf}, 
we now know many
properties of  one-dimensional manifolds (directed polymers)
 in random media. There have also been numerous
numerical simulations; some recent studies can be found in 
Ref.~\cite{mezard,kim,hhk}.

Recently, strong flux pinning effects~\cite{civale,koncz,budhani}
 exhibited by samples of high temperature superconductor with 
{\it extended} defects such as columnar faults 
and twin planes lead naturally
to the investigation of competition between extended 
and point defects~\cite{hnv,balents}.
It has been argued in the case of many interacting flux lines that pinning
by extended defects are {\it weakened} by point defects~\cite{hnv}.
There have also been a number of studies on the competing effects between
extended and point defects on a {\it single} 
flux line or a directed polymer~\cite{mk,zhh,tl,bk1,kolo1,kolo2,bk2,kolo3}.
An early study of this type was done by Kardar~\cite{mk}, 
who investigated numerically
the pinning of a directed polymer to a {\it line defect} in the presence of a
background of point defects in 1+1 dimensions. The polymer was
found to depin from the line defect, if the pinning potential of the
line defect is smaller than a certain threshold value. Critical behavior
associated with the depinning transition was later 
investigated in more detail by
Zapatocky and Halpin-Healy~\cite{zhh}. The results of Refs.~\cite{mk} and
\cite{zhh} have been challenged by Tang and Lyuksyutov~\cite{tl}, 
who argued against the existence of a depinning transition in two dimensions,
based on large scale simulations and
approximate renormalization-group analysis on a hierarchical lattice.
Tang and Lyuksyutov proposed instead that a directed polymer is
{\it always} localized, although weakly, to the line defect
in 1+1 dimensions, and that the depinning transition only
exists above 1+1 dimensions. This conclusion is supported by a recent
 study of Balents and Kardar~\cite{bk2}, who performed numerical simulations,
and also developed a functional renormalization group analysis by
describing the directed
polymer as a generalized $D$-dimensional manifold and studying 
the depinning transition using a {\it double} expansion (in the
dimensionality of the manifold $D=4-\delta$ and the dimensionality
of the embedding space $d = d_c(D)+\epsilon$.) On the other hand,  analysis of the directed polymer itself by Kolomeisky and Straley have led to different conclusions when different renormalization group  Ansatz were 
used~\cite{kolo1,kolo2,kolo3}.

In this paper, we give a {\it systematic} analysis of the competition
in pinning  between
point and line defects for a directed polymer. By exploiting 
known knowledge
of the randomly pinned directed polymer in 1+1 dimensions
 in the absence of any extended
defects~\cite{hf}, we construct a renormalization-group analysis {\it directly}
in 1+1 dimensions,  the critical dimension for this problem. 
Our results prove the conclusion of Ref.~\cite{tl}, that the
polymer is always pinned at and below 1+1 dimensions, and that the
depinning transition only occurs above two dimensions. The existence
of weak localization at the critical dimension is understood in terms
of the anomolous large scale excitations of the directed polymer,  known as the
``droplet excitations"~\cite{hf}. Critical scaling behaviors 
at the depinning transition are then obtained using 
a $1+1+\epsilon$ dimensional expansion. 

The paper is organized as follows: We first introduce the known properties
of the randomly pinned directed polymer in Section II.
We consider the effect of pinning by a line defect using phenomenological
scaling arguments, which establishes 1+1 dimensions to be the critical
dimension.  In Section III, we attempt to solve the problem using
the replica method at the critical dimension. Despite the existence of 
an exact solution via the Bethe Ansatz~\cite{bethe,p} when the extended defect is absent, we failed to develop a systematic and controlled way of incorporating a weak extended defect. We next use an uncontrolled Hartree
approximation, which contains the right physical ingredients, but
 overestimates the effect
of point disorder. A {\it systematic}
 analysis taking advantage of known mapping
of the directed polymer to the noisy-Burgers' equation~\cite{hf}
 is presented in Section IV. Through explicit calculations, facilitated by the 
use of the mode coupling approximation in $1+1$ dimensions, we construct
the renormalization-group recursion relation and derive various
scaling behaviors near the critical dimension.  We show that the results
can be understood naturally in terms of the anomolous droplet 
excitations~\cite{hf}.
In Section V, we generalize our method to describe the directed polymer 
pinned by  a variety of
extended and point defects, including columnar pin of extended
range and trajectory, and point defects with long range correlations. 
Some useful relations about the randomly pinned
directed polymer, in particular, the mode coupling approximation,
 is summarized in Appendix A, and a number of detail
calculations are relegated to Appendix B.

\newpage
\section{Phenomenological scaling analysis}

We describe the statistical mechanics  of a randomly pinned directed polymer
of length $t$ by the Hamiltonian
\begin{equation}
{\cal H}_0 = \int^t_0 dz\left\{\frac{\kappa}{2}\left(\frac{d\bxi}
{dz}\right)^2 + \eta[\bxi(z),z]\right\}, \label{H0}
\end{equation}
where $\bxi(z)\in \Re^{d_\perp}$ denotes the transverse displacement 
of the polymer in $d = d_\perp + 1$ embedding dimensions (with
$d_\perp$ being the co-dimension),
$\kappa$ is the polymer line tension,
and  $\eta(\bx,z)$ describes a background medium of uncorrelated
point defects. The random potential
is assumed to be Gaussian distributed
with mean zero and the variance
\begin{equation}
\overline{\eta(\bx,z)\eta(\bx',z')} 
= 2\Delta\delta^{d_\perp}(\bx - \bx')\delta(z-z'), \label{eta}
\end{equation}
where the overbar denotes disorder average.

The randomly pinned directed polymer has been
the topic of detailed investigations during the last
decade~\cite{cayley}--\cite{hhk}. 
The emerging qualitative picture for a long polymer ($t\to\infty$) is the 
following: At low temperatures, the 
static properties of the polymer are dominated by the random potential
 and are
controlled by one or a few optimal path(s) which minimize the total free
energy~\cite{fh,hf}. 
To take advantage of fluctuations in the random potential $\eta$,
the optimal path executes large transverse wandering 
[solid line of Fig.~1(a)]. If we fix one
end of the polymer, then the root-mean-square displacement
of  the other end of an optimal path is 
\begin{equation}
X(t) = B t^\zeta, \label{Xrms}
\end{equation}
where $3/4 > \zeta >1/2$ is an 
universal exponent which depends on $d$. The result
$\zeta =2/3$ in $d=1+1$ has been obtained exactly by 
a number of methods~\cite{bethe,hhf}.
Numerical calculations give $\zeta \approx 5/8$ in $d=2+1$.
Displacement of the fixed end of the polymer by a distance $r \ll X(t)$
typically result in a 
rearrangement of the optimal path within a segment of length
$\tau \sim (r/B)^{1/\zeta}$ from the fixed end, as shown in Fig.~1(a).
For sufficiently large displacement, i.e., 
for $ r \gg B t^{\zeta}$, the optimal path is completely
changed [dotted line of Fig.~1(a)], 
with a completely different free energy [Fig.~1(b)]. The typical 
free energy difference $\Delta F_0$ between such independent paths is
\begin{equation}
\Delta F_0 \approx At^{\theta}, \label{Frms}
\end{equation}
with an identity 
\begin{equation}
\theta=2\zeta-1 \label{id}
\end{equation}  relating the two exponents.
$\Delta F_0$ also sets the scale of the free energy difference
between the optimal path and a typical path.

Thermal fluctuations do not change this picture
qualitatively. They merely wipe out the fine structure of 
the random potential, leading  
to temperature dependent amplitudes. For instance, 
$B \propto (\Delta/\kappa T)^{1/3}$ and $A \propto \kappa B^2$ in $d=1+1$.
In $d \le 2+1$ dimensions, this picture is  in fact 
correct for all temperatures. In $d > 2+1$, the polymer undergoes a continuous
phase transition and becomes dominated by thermal fluctuations rather
than disorder at sufficiently high temperatures.
We shall however be focused on the more interesting low temperature phase 
throughout this paper.

Let us consider the influence of an additional 
pinning potential, $U_p(\bx)$, in the form of a line defect with a short range
$a$, located at the origin. We model the defect by the Hamiltonian
\begin{equation}
{\cal H}_1 = \int^t_0 dz \  U_p[\bxi(z)], \label{H1}
\end{equation}
with $U_p(\bx) = - U$ for $|\bx| < a$, and $U_p(\bx) = 0$ for $|\bx| > a$.
Clearly, such an attractive potential favors configurations of the polymer
close to the origin, i.e., it attempts to {\it localize} the polymer.
On the other hand, the polymer will try to wander {\it away}
from the defect in order to take advantage of  fluctuations 
in the background random potential $\eta(\bx,z)$. Depending on the outcome
of this competition, the pinning potential
$U_p$ may have a number of possible effects:
It may always or never localize the polymer for any finite
pinning strengths $U$, or it may localize the polymer only above a 
certain critical pinning strength $U_c$, thus giving rise to a depinning
transition.

To gain an understanding of the competing effects,
we first consider the pure problem with $\eta(\bx,z) = 0$.
Let 
\begin{equation}
W^{(0)}(\bx,t) = \int^{(\bx,t)}_{(0,0)} {\cal D}[\bxi] 
e^{-({\cal H}_0^{(0)} + {\cal H}_1)/T}
\end{equation}
be the Boltzmann weight of propagating the polymer from $(0,0)$ to $(\bx,t)$,
with ${\cal H}_0^{(0)} = {\cal H}_0[\eta=0]$ and the superscript $(0)$
indicating the absence of the random potential,
then $W^{(0)}$ satisfies the diffusion equation
\begin{equation}
T \frac{\partial}{\partial t} W^{(0)}(\bx,t) 
= \frac{T^2}{2\kappa}\nabla^2 W^{(0)} - U_p(\bx)W^{(0)}(\bx,t).
\label{Seqn.0}
\end{equation}
Our problem is equivalent to that of a $d_\perp$-dimensional
(imaginary-time)
quantum mechanical particle with a pinning well $U_p$ at the origin. 
In the thermodynamic limit $t \to \infty$, 
the statistical mechanics of the polymer
is given by the ground state of the quantum problem whose solution is
well known: There is always a bound state
for arbitrarily weak pinning potential in $d_\perp \le 2$, where the polymer
is always bounded to the line defect.
For $d_\perp > 2$, a critical pinning strength of the order 
$U^{(0)}_0 \equiv T^2/(2\kappa a^2)$ is necessary to have a  bound state.
A depinning transition occurs at $U = U^{(0)}_c \sim U^{(0)}_0$.
Fig.~2 summarizes the phase diagram of the pure system in various dimensions.

It is useful to understand these results by simple physical
considerations: Assume first that the polymer is completely localized
within the pinning well. Then the energy gained (compared to 
a free polymer) is $U$ per length, and the entropy cost (or the 
kinetic energy cost of localizing a quantum  particle)
is of the order $U_0^{(0)}$ per length. Thus if $U \gg U_0^{(0)}$,
the total free energy of the polymer is $ F^{(0)} = -U + U_0^{(0)} < 0$, and
the polymer is localized. 
However, if $U \lesssim U_0^{(0)}$, then $F^{(0)} >0$ and
the polymer will not be completely localized to the well since
the free energy exceeds that of a free polymer in this case. 
An alternative scenario when $U \lesssim U_0^{(0)}$ is 
to have the polymer completely delocalized, ignoring the
existence of the weak pinning potential. This is self-consistent 
for $d_\perp>2$, since for a polymer with one end fixed at the origin, the 
accumulated probability of a long free polymer returning to the origin is 
$t^{(2-d_\perp)/2} \to 0$. Thus we see that for $d_\perp >2$, the polymer is
essentially free for $U \ll U_0^{(0)}$, and is completely localized if $U \gg
U_0^{(0)}$. The simplest phase diagram is then to have 
a depinning transition separating the
pinned and free phases, occuring at $U = U_c^{(0)} \sim U_0^{(0)}$.
On the other hand,
for $d_\perp \le 2$, the probability of a long free polymer returning 
to the origin is of order $1$. Thus the completely delocalized phase
is not self consistent here,
and the effect of a weak pinning potential can never be ignored in the limit
$t \to \infty$.

The above can be stated more quantitatively by comparing the average pinning
energy $\langle {\cal H}_1 \rangle^{(0)}$ with the available 
thermal energy $T$. A useful parameter to focus on is the dimensionless ratio
$g^{(0)} = \langle {\cal H}_1 \rangle^{(0)} / T$.
The polymer is free if $g^{(0)} \ll 1$ and is completely localized 
if $g^{(0)} \gg 1$ in the limit $t \to \infty$.
If we just perform the thermal average $\langle \ldots \rangle^{(0)}$ by the
Wiener measure of a {\it free} polymer, $e^{-{\cal H}_0^{(0)}/T}$, 
then we obtain
\begin{equation}
g^{(0)}_0 = \frac{1}{\epsilon^{(0)}}\left[\left(\frac{t}{a_\|^{(0)}}\right)
^{\epsilon^{(0)}} - 1 \right]
\frac{U}{U_0^{(0)}} \label{g00}
\end{equation}
where $\epsilon^{(0)} = (2- d_\perp)/2$ and $a_\|^{(0)} = (\kappa/T) a^2$ 
is the short distance cutoff along the length of the polymer.
Thus the free polymer phase is obtained for $d_\perp > 2$ 
(i.e., $\epsilon^{(0)} < 0 $) if $ U < U_c^{(0)} \propto
|\epsilon^{(0)}| U_0^{(0)}$.  
For $U > U_c^{(0)}$, the polymer is free only for
\begin{equation}
t < l_\parallel^{(0)} = a_\|^{(0)} \left[\frac{U}{U_c^{(0)}}-1\right]
^{-\frac{1}{|\epsilon^{(0)}|}}.
\end{equation}
$l_\|^{(0)}$ is a crossover length above which 
the polymer becomes localized to the
pinning potential. From it, we can define a localization length
\begin{equation}
l_\perp^{(0)} =\left[\frac{T}{\kappa} l_\|^{(0)}\right]^{1/2} 
=  a \left[\frac{U}{U_c^{(0)}}-1\right]^{-\frac{1}{2|\epsilon^{(0)}|}},
\qquad d_\perp  > 2\label{L0.1}
\end{equation}
which characterize the typical transverse excursion of the polymer.
On the other hand, for $d_\perp < 2$ where $\epsilon^{(0)} > 0$, 
$g^{(0)}_0 \gg 1$  for arbitrary
small $U$ in the limit of long polymer $t \to \infty$. 
Defining the crossover
and localization lengths in the same way, we have
\begin{equation}
l_\perp^{(0)}  =  a [U/U_c^{(0)}]^{-\frac{1}{2|\epsilon^{(0)}|}}, 
\qquad \epsilon^{(0)} > 0 \label{L0.2}
\end{equation}
with the polymer delocalized only in the limit $U \to 0$.
Qualitatively similar behavior is obtained in $d_\perp = 2$, with 
\begin{equation}
l_\perp^{(0)}  =  a e^{U_c^{(0)}/U}. \label{L0.3}
\end{equation}
Eqs.~(\ref{L0.1}) -- (\ref{L0.3}) are valid as long as $l_\perp^{(0)} \gg
a$. Throughout this paper, we shall only be interested in this critical regime,
 where the scaling behavior of localization length is insensitive to
the detail shape  of $U_p$.

Let us now return to the problem of a {\it randomly pinned}
 directed polymer,  i.e., with $\eta(\bx,z) \ne 0$,
and estimate  the competition between the pinning
and the random potential that occurs in this case.
Just as the localization of a polymer costs entropy in the pure problem, here,
localization prevents the polymer from seeking out favorable regions of the
random potential $\eta$ far from the origin, and therefore leads to
a {\it loss } in the random component of the  energy that could otherwise
be gained even as $T\to 0$.
If a polymer is localized within a distance $l_\perp$ about the origin,
then it is consisted of a number of uncorrelated segments of length
$l_\| \approx (l_\perp/B)^{1/\zeta}$, each of which having a free energy
of order $Al_\|^\theta$ higher than that of the delocalized, optimal 
path (see Fig.~3). Thus the free energy cost of localization is of the order
$ A (l_\perp/B)^{(\theta-1)/\zeta}$ per length. 
This plays the role of the entropy cost in the localization of a free polymer.
For a very strong pinning potential, the polymer is again always 
localized completely within the  pinning well because the energy per length
gained, $U$, always exceeds  the random energy cost of localization, 
which is of the order $U_0 = A (a/B)^{(\theta-1)/\zeta}$ per length.

The effect of a weak pinning potential can be estimated perturbatively
from $\langle {\cal H}_1 \rangle$ as just described  for the pure problem,
except that it now depends on the realization of the random potential $\eta$.
The natural quantity to examine is the average energy gained by the polymer
in the presence of the pinning potential,
\begin{equation}
\delta F = \overline{ \langle {\cal H}_1 \rangle },
\end{equation}
given that one end is fixed at the origin (as shown in Fig.~3).
As in the case of the pure problem, this is just the
average of the accumulated return probability of the polymer to the origin.
Since the rms displacement is $B t^\zeta$ in the absence of the pinning
potential $U_p$, 
then $\delta F_0 \propto U (a/Bt^\zeta)^{d_\perp}t$
to leading order in $U$. To determine the effect of the pinning potential, 
it is necessary to compare $\delta F_0$
to $\Delta F_0 \approx A t^\theta$, the intrinsic variations in the free
energy [see Fig.~1(b)].  The dimensionless ratio that characterizes
the strength of pinning is now
\begin{equation}
g_0 = \frac{\delta F_0}{\Delta F_0} 
\propto \frac{U}{U_0} \left(\frac{t}{a_\|}\right)^\epsilon,
\end{equation} 
to leading order in $U/U_0$, with
$a_\| =  (a/B)^{1/ \zeta}$ being the short distance cutoff along the length 
of the polymer, and 
\begin{equation}
\epsilon(d) = 1 - d_\perp\zeta - \theta = 2 - (d+1) \zeta(d). \label{eps}
\end{equation}
The critical dimension is $d=1+1$ since $\zeta(2) = 2/3$
exactly~\cite{bethe,hhf}.
If $g_0 \ll 1$, which is the case when $\epsilon < 0$,
then the weak pinning
potential is irrelevant in the thermodynamic limit. 
If $\epsilon > 0$, then $g_0 \gg 1$ and  even  a small $U$
will completely change 
the energy landscape of the randomly pinned directed polymer. For example,
the free energy of the pinned state in Fig.~3 now becomes much lower
 than that of the optimal path of the unperturbed system.
The apparent difference between the problem
with point disorder and the pure problem is the behavior of $g_0$
at the critical dimension
where $\epsilon = 0$: For the pure problem,
$g^{(0)}_0$ diverges logarithmically according to Eq.~(\ref{g00}), 
indicating the marginal {\it relevance}
of the pinning potential.  But with point disorder, 
\begin{equation}
g_0 \propto U/U_0 \propto U a / (A B) \label{g0}
\end{equation}
 remains finite. This  naively suggests the {\it irrelevance} of 
a small pinning potential at the critical dimension. If this is true,
then there will have to be a depinning transition at the critical dimension
$d=1+1$, since a strong enough pinning potential still produces a localized
state.  The simplest phase diagram in this case [Fig.~4(a)] will 
look quite different
from that of the pure problem (Fig.~2). Fig.~4(a)
is actually consistent with early numerical results of Kardar~\cite{mk} and
Zapatocky and Halpin-Healy~\cite{zhh}, 
who find a depinning transition in $d=1+1$. However, through a systematic renormalization-group analysis presented in Sec.~IV, 
we will find that $U$ itself is renormalized and
diverges logarithmically, while all other parameters such as $A$ and $B$
only suffer finite renormalization. These results lead to a logarithmically
 diverging $g$ and hence
the marginal {\it relevance} of pinning at the critical dimension.
Thus the phase diagram for the randomly pinned directed polymer is like
that of the pure problem (Fig.~2), but with a shifted critical dimension and
$\epsilon$, as shown in Fig.~4(b).  This result is supported by the
recent large scale numerical studies of Refs.~\cite{tl,bk2}.
Renormalization group analysis can then be used to obtain the critical
behavior at the depinning transition, in particular, the divergence of the
correlation length,
\begin{equation}
l_\perp = a\left(\frac{U}{U_c} - 1\right)^{\nu_\perp},
\end{equation}
where $\nu_\perp$ is the liberation exponent.

\newpage
\section{Analysis in Replica space}

A convenient way to treat one dimensional objects like the directed
polymer is the transfer matrix approach described in Sec.~II.  The {\it full}
Boltzmann weight (or the restricted partition function) $W(\bx,t)$ of a 
polymer propagating
from $({\bf 0},0)$ to $(\bx,t)$ in a random medium $\eta(\bx,t)$
 is described by an equation analogous
to Eq.~(\ref{Seqn.0}), except with $U_p(\bx)$  replaced by $U_p(\bx)+\eta(\bx,t)$. 
All physical properties follow from disorder averages of the free energy
$\overline{F(\bx,t)}=-T \overline{\ln W(\bx,t)}$ and derivatives thereof.
One approach to performing this average is the replica method, which
exploits the identity 
$\ln W(\bx,t)=\lim_{n\to 0}\frac{1}{n}(W^n(\bx,t)-1)$.
Up to an exchange of the $n\to 0$ and the thermodynamic limit (see below),
$\overline{F(\bx,t)}$ is given by the $n^{th}$ moment of $W$,
$\overline{W^n(\bx,t)}\equiv {\cal W}(\bx,...\bx ;t)$, where
\begin{equation}
{\cal W}(\bx _1,..., \bx_n;t)=\prod^n_{\alpha=1}
\int^{(\bx_{\alpha},t)}
_{({\bf 0},0)}
{\cal D}[\bxi_{\alpha}]e^{-{{\cal H}_n}/T}, \label{Wn}
\end{equation}
$\alpha \in \{1, \ldots, n\}$ is the replica index, and
 ${\cal H}_n$ denotes the replica Hamiltonian
\begin{equation}
{\cal H}_n = \sum ^n_{\alpha=1} \int^t_0 dz
\left\{\frac{\kappa}{2}\left(\frac{d\bxi_\alpha}{dz}\right)^2 
+ U_p[\bxi_{\alpha}(z)]-\sum_{\beta\neq \alpha}\frac{\Delta}{T}
\delta^{d_\perp} [\bxi_{\alpha}(z)-\bxi_{\beta}(z)] \right\}. \label{Hn}
\end{equation}
For analytic simplicity, we approximate the columnar pinning potential
by a $\delta$-function, i.e., $U_p(\bx) = - u \delta^{d_\perp}(\bx)$ with 
$u \approx U a^{d_\perp}$, from here on.
Applying the transfer matrix approach directly to Eq.~(\ref{Wn})
 leads to the  Schr\"odinger-like equation
\begin{equation}
T\frac{\partial}{\partial t}{\cal W} (\bx _1,...,\bx _n;t)
=-\hat{\cal H }(n)\  {\cal W}(\bx_1,...,\bx_n;t), \label{Seqn}
\end{equation}
where 
\begin{equation}
\hat{\cal H}(n)=\sum^n_{\alpha=1}
\left\{-\frac{T^2}{2\kappa}\grad^2_\alpha
-u\delta^{d_\perp}(\bx _{\alpha})-\sum_{\beta\not=\alpha}
\frac{\Delta}{T}\delta^{d_\perp}(\bx_ {\alpha}-\bx _{\beta})\right\},
\label{hartree.H}
\end{equation}

As usual, Eq.~(\ref{Seqn}) can be solved with the Ansatz
\begin{equation}
{\cal W}(\bx _1,..., \bx _n ;t)=\sum ^{\infty}_{j=0} a_j e^{-E_j(n) t/T}
\Phi_j(\bx _1,...,\bx _n)
\end{equation}
and $\hat{\cal H}(n) \Phi _j = E_j(n) \Phi _j$. 
In $d_\perp = 1$ and for $U_p = 0$, the ground state 
wave function is easily found by using the Bethe-Ansatz~\cite{bethe},
 $\Phi_0(x_1,...,x_n) \propto \exp\{ - \frac{\Delta \kappa}{T^3}
\sum_{\alpha < \beta}|x_\alpha - x_\beta| \}$. 
A similar solution follows for the ground state of a randomly pinned
directed polymer confined to the semi-infinite plane $x \ge 0$ and
 subject to an attractive potential $U_p$ at $x=0$~\cite{bethe}. 
Upon changing the strength of the
attractive potential, one obtains a depinning transition for the
half-plane problem when
 the ground state energy  becomes larger than that for $U_p =0$.
However, searches for a similar solution to the Hamiltonian (\ref{hartree.H})
have not been successful.
A naive application of the perturbation theory for small $U_p$ 
starting from 
the wave function  $\Phi_0(x_1,...,x_n)$  fails likewise, 
because   the low energy {\it excited} states of $\hat{\cal H}(n)$
are not uniquely defined in the limit
the number of particles $n \to 0$~\cite{p}. 
This ambiguity, arising from  the exchange of the $n\to 0$ and the
thermodynamic limits, is a well-known ``trouble spot'' for
the replica method, and often calls for elaborate schemes with
broken replica symmetry~\cite{mp}.
 For the problem at hand, the ground
state itself does not involve replica symmetry breaking~\cite{bethe,p}. 
However, it is demonstrated
in Ref.~\cite{p} that states with broken replica symmetry could have
energies  arbitrary close to the ground state energy $E_0(n)$.
These will dominate in any perturbative calculations.

We shall circumvent the problem of replica symmetry breaking 
by introducing a completely different method in Sec.~IV. For now,
we consider another limit, $\Delta \to 0$, where the system  almost
decouples into $n$ one-particle problems. Here an attractive potential 
$U_p$ always gives a bounded ground state in $d_\perp = 1$.
In term of the	 reduced length $r = x/x_0$ where $x_0=T^2/(2u\kappa)$,
the reduced Hamiltonian 
$\hat{h}(n) = \hat{\cal H}(n)/\frac{T^2}{2\kappa x_0^2}$  becomes
\begin{equation}
\hat{h}(n) = - \sum_\alpha \left\{ \frac{\partial^2}{\partial r_\alpha^2} + 2
\delta(r_\alpha) + \frac{2\Delta}{uT}\sum_{\beta\neq \alpha}
\delta(r_\alpha - r_\beta) \right\}. \label{h.hat}
\end{equation}
We now attempt to treat the case of {\it weak} inter-replica interaction
$\Delta \ll uT$ by using the Hartree approximation~\cite{no}.
Making the Ansatz for the wave function 
$\Phi(r_1, r_2, \ldots, r_n) = \phi_1(r_1)\phi_2(r_2)...\phi_n(r _n)$
and minimizing the reduced energy ${\cal E}(n) = 
\int dr_1 \ldots dr_n\Phi\hat{h}(n) \Phi$
under the conditions 
$\int dr \phi^2_{\alpha}(r) = 1$ which we impose by  Lagrangian multipliers 
$\eps_{\alpha}$, we get a set of 
Hartree equations~\cite{no}. These simplify to a single
equation if we impose symmetry between the replicas 
$\phi_{\alpha}(r)=\phi(r)$ and $\eps_\alpha = \eps$ for all $\alpha$'s,
\begin{equation}
\left[ \frac{\partial^2}{\partial r^2} +2 \delta(r)+4(n-1)g^{-1}_0
{\phi}^2(r)-\eps \right] \phi (r) = 0,\label{hartree.eqn}
\end{equation}
where 
\begin{equation}
 g_0^{-1}= \Delta/(uT). \label{g0.2}
\end{equation}
This equation can be solved easily with the boundary
 condition  $\phi(r) = \phi'(r) = 0 $
for $|r| \to \infty$, giving  a localized wave function
\begin{equation}
\phi(r) = e^{-r \sqrt{\eps}}\left[\frac{\sqrt{\eps}+\sqrt{\eps 
-2(n-1)g_0^{-1}\phi^2(r)}}{\sqrt{1+\eps}}\right] \label{hartree.1}
\end{equation}
with a negative ground state energy
\begin{equation}
{\cal E}_0(n)=-\frac{n}{3}(1+\sqrt{\eps}+\eps)
= -n\left[1+(n-1)g_0^{-1}+\frac{1}{3}(n-1)^2g^{-2}_0\right] 
\label{hartree.2}
\end{equation}
where $\eps =[1 + (n-1)g_0^{-1}]^2$, and 
the localization length  is 
\begin{equation}
\l_{\perp} =x_0/\sqrt{\eps}
\propto \left( 1+ (n-1)g_0^{-1}\right)^{-1}. \label{hartree.3}
\end{equation}
Note that the Hartree results Eqs.~(\ref{hartree.1}) -- (\ref{hartree.3})
 are in fact the {\it exact} solutions of the 
Hamiltonian (\ref{hartree.H}) in the limit $g_0^{-1} \to  0$,
where the very strong pinning potential always 
localizes the directed polymer within $l_\perp$ of the origin.
However, we see that in the limit $n\to 0$, the random potential
characterized by $g_0^{-1}$ tends to {\it increase} the localization
length.
If we extrapolate the Hartree result Eq.~(\ref{hartree.3}) to finite
value of $g_0^{-1}$, then 
we find an instability, i.e., $l_\perp \to \infty$ as $g_0^{-1} \to 1^{-}$.
An obvious interpretation of this instability is the occurrence 
of a depinning transition.

On the other hand, an analogous Hartree
calculation for the problem of an attractive potential in the half
plane yields again the solution (\ref{hartree.1}), 
except with $\eps = (1+2(n-1)g_0^{-1})^2$. This result suggests 
a delocalization transition at $g_{c}^{-1} = 1/2$ with 
$l_\perp \propto (g_0 - g_c)^{-1}$, or $\nu _\perp =1$ for the half-plane
problem. However, the exact Bethe-Ansatz calculation~\cite{bethe}
gave $g_{c}^{-1} =1$ and $\nu _\perp =2$. Thus we see that while
the Hartree solution gives the qualitative effect of a
background random potential --- a tendency to {\it depin} the directed
polymer from the origin --- it {\it overestimates} the influence of
randomness and cannot be used reliably to characterize
 the depinning transition quantitatively.
Indeed, we will show in the following section  that 
the directed polymer is always pinned to the origin 
for {\it all} finite values of
 $g_0$ in $d_\perp = 1$ dimensions.

[ Note: Although our Hartree equation (\ref{hartree.eqn})
 agrees with that considered previously by Zhang~\cite{zhang}
 if we set $u=0$, 
the physics of our solution is completely different:
In the limit $n \to 0$ which we consider, 
the disorder leads to a {\it repulsive}
interaction $g_0^{-1}\delta(x_\alpha - x_\beta)$
 between the replicas, which makes a change of the roughness 
exponent $\zeta$ from $1/2$  for free polymers 
to the larger value $2/3$ possible. 
It is the pinning  potential $U_p$ that confines the solution 
close to the origin. In contrast, Zhang considered the {\it large}-$n$ 
limit where the interaction between the replicas is {\it attractive}, 
leading to the unphysical solution of an
exponentially decreasing wave function even if $u=0$. ]

\vfill
\newpage
\section{The Renormalization-Group Analysis}

As we have seen in the Sections II and III, both naive scaling arguments
and the uncontrolled Hartree solution of the replicated system suggest
that the polymer undergoes a depinning transition in the critical dimension
$d=1+1$, consistent with the phase diagram sketched in Fig.~4(a). However,
we have not been able to construct a {\it controlled}
perturbative study for weak pinning potential $U_p$ within 
the replica formalism, despite the knowledge of the exact ground state.
The technical problem encountered is the occurrence of replica symmetry
breaking for the low energy excited states.
In this section, we shall formulate
 a perturbative renormalization group study of the depinning transition
in $1+1$ dimensions without ever introducing the notion of replica.
We shall take advantage of the mapping of the directed polymer 
to the hydrodynamics  of the noisy-Burgers' equation~\cite{hhf,iv,hf}. 
Many details of the mapping have been 
discussed in Ref.~\cite{hf} and some are summarized in Appendix A. 
Here, we will outline our approach and quote
 the results. A number of detailed calculations are given in Appendix B.

\subsection{Formalism}

It is useful to introduce first a formal language to characterize the
properties of the directed polymer in the absence of 
the pinning potential $U_p$.  For convenience, we shall fix
one end $\bxi(t)$ of the polymer at some arbitrary point $\bx$.
This is implemented by introducing the one-point restricted partition
function,
\begin{equation}
Z_0(\bx,t) = \int {\cal D}[\bxi] \  \delta^{d_\perp}(\bxi(t)-\bx) 
e^{-{\cal H}_0/T}, \label{Z0}
\end{equation}
where the subscript $0$ is used to indicate $U_p = 0$.
All thermal averages taken with respect to $Z_0(\bx,t)$
will be denoted by $\langle \ldots \rangle_{\bx,t}$.  
The disorder averaged probability
of finding a segment $\bxi(t_0)$ of the polymer at position $\by$
given that the other end is fixed at $(\bx,t)$ is
\begin{equation}
G^{(1,1)}_t(\bx-\by,t-t_0) \equiv \overline{\langle 
\delta^{d_\perp}(\bxi(t_0)-\by) \rangle}_{\bx,t}, \label{G}
\end{equation}
where we use the subscript $t$ to emphasize the dependence of $G^{(1,1)}$
on the total polymer length $t$. 
The functional form of $G^{(1,1)}$ has been investigated in details
elsewhere~\cite{hf} and is summarized in Appendix A. Here, we
just mention the qualitative behavior,  
$G_t^{(1,1)}(\br,\tau) = (B \tau^\zeta)^{-d_\perp}$ for
$r \ll B \tau^\zeta$ and vanishes rapidly as $r \gg B\tau^\zeta$.

It is also useful to consider the free energy of the one-point restricted
polymer, 
\begin{equation}
F_0(\bx,t) = - T \log Z_0(\bx,t),
\end{equation}
which satisfies the noisy-Burgers' equation~\cite{fns,kpz,medina}
\begin{equation}
\partial_t F_0(\bx,t) = \frac{T}{2\kappa}\grad^2 F_0 - \frac{1}{2\kappa} (\grad
F_0)^2 + \eta(\bx,t). \label{kpz.eqn}
\end{equation}
The free energy correlation function is 
\begin{equation}
C_t(\bx-\by) = \overline{\left[F_0(\bx,t)-F_0(\by,t)\right]^2}, \label{C}
\end{equation}
with 
\begin{eqnarray}
&C_t(\br) = 2 A^2 t^{2\theta} \qquad &{\rm for} \qquad r \gg Bt^\zeta,\label{ct}\\
&C_t(\br) \propto A^2(r/B)^{2\theta/\zeta} \qquad 
&{\rm for} \qquad r \ll Bt^\zeta.\label{cr}
\end{eqnarray}
The sample-to-sample free energy variation introduced in Eq.~(\ref{Frms})
is just 
\begin{equation}
\Delta F_0 = \sqrt{C_t(\infty)/2} = At^\theta.
\end{equation}
Again, the scaling form for $C_t$ is summarized in Appendix A.
As we shall see, the functions $G$, $C$, together with
a number of other distribution functions, will allow us to
compute the effect of an additional pinning potential perturbatively.

In terms of $Z_0$, the full partition function $Z(\bx,t)$ of the polymer 
in the presence of the pinning potential 
$U_p(\bxi(z)) = - u  \delta^{d_\perp}(\bxi(z))$  is just
\begin{equation}
Z(\bx,t) = Z_0(\bx,t) \langle e^{-{\cal H}_1/T} \rangle_{\bx,t}, \label{Z}
\end{equation}
where ${\cal H}_1 = \int_0^t dz \ U_p$ as in Eq.~(\ref{H1}). For convenience,
 we again use the continuum delta function 
supplemented by a short distance cutoff $a$ 
(with $u=Ua^{d_\perp}$) to model the pin.

\subsection{The Renormalization Group Analysis}

We now investigate the effect of ${\cal H}_1$ on the ``bare" system $Z_0$
perturbatively. Clearly the analysis is complicated by the fact that 
the ``bare" system itself is glassy and thus highly nontrivial.
However, to understand whether the phase diagram belongs to that of
Fig.~4(a) or Fig.~4(b), we only need to compute the marginal relevancy of
the pinning potential in $1+1$ dimensions.  Fortunately, 
the randomly pinned directed polymer in $d=1+1$ is  one of the
very few glassy systems for which a great deal is known.
In particular we note that there are two pieces of exact information:
(i) The disorder averaged thermal displacement is the same as that of 
the pure system, i.e.,
\begin{equation}
\overline{ \langle [\xi(t) - \xi(0)]^2 \rangle_{x,t}^c } 
= \frac{T}{\kappa} t,\label{gi} 
\end{equation}
due to a statistical tilt symmetry~\cite{hf,schulz}
 preserved by the point disorder.
(ii) A fluctuation-dissipation theorem for the noisy-Burgers'
equation~\cite{fns,medina} which states that
\begin{equation}
\overline{\partial_x F_0(x,t)\partial_y F_0(y,t)} = \frac{\Delta \kappa}
{T} \delta (x-y) \label{fdt}
\end{equation}
in the limit of large $t$ (see Appendix A).
Eq.~(\ref{fdt}) implies that $C_t(r) = (\Delta\kappa/T) |r|$.
Comparison with Eq.~(\ref{cr}) yields $2\theta/\zeta= 1$, and 
$A^2/B \propto \Delta\kappa/T$. Along with the exponent identity
Eq.~(\ref{id}), we have $\theta = 1/3$, $\zeta=2/3$, and dimensional
analysis gives $B\propto (\Delta/\kappa T)^{1/3}$.
We see that our ``bare'' problem,  the fixed point 
of the randomly pinned directed polymer, 
 is  a fixed plane spanned by the axis $\kappa$ and $\Delta\kappa/T$,
or alternatively by the parameters $A\propto \kappa B^2$ and $B$.
Hence all disorder averaged functions depend {\it only} on these two 
parameters in the limit of large $t$. For example,
the bare coupling constant of the problem is defined as
 $g_0 = \delta F_0/\Delta F_0$ where $\Delta F_0 = A t^{1/3}$ and
\begin{eqnarray}
&\delta F_0 & = \overline{\langle{\cal H}_1\rangle_{0,t}} = -u \int_0^t dz
\overline{\langle \delta[\xi(z)] \rangle_{0,t}} \nonumber \\
& & = -u \int_0^t dz \ G_t^{(1,1)}(0,t-z). \label{dF}
\end{eqnarray}
Scaling form for $G^{(1,1)}$ (see Appendix A) 
yields $\delta F_0 \propto -(u/B)t^{1/3}$, hence $g_0 \propto u/(AB)
\propto uT/\Delta$ as in Eqs.~(\ref{g0}) and (\ref{g0.2}). 
In subsequent calculations, we shall use $g_0 \equiv
u/(\kappa B^3)$.
Following standard renormalization group treatment, we now consider
the renormalization of the parameters $\kappa$, $B$, and $u$
  due to the perturbation ${\cal H}_1$.
[Note that  the effect of the pinning potential will obviously depend on
the position of the fixed end $\xi(t)$. If $\xi(t)$ is sufficiently
far away from the origin, the polymer will never feel the effect of $U_p$.
In our calculations, we shall fix $\xi(t)$ at the origin (as shown in Fig.~3)
to evaluate the maximal effect of the pinning potential.]

We begin with the renormalization of the stiffness $\kappa$, since the
presence of a pinning potential breaks the statistical tilt symmetry and
hence the exact relation Eq.~(\ref{gi}). Adding a 
term $- h \int_0^t dz (d\xi/dz)$ 
to ${\cal H}$ and using the transformation $\xi (z) 
\rightarrow \xi (z) + h (t-z)/ \kappa  $
and the property  $\eta(\xi (z) + h(t-z) / \kappa,z) = \eta (\xi(z),z)$
in  the statistical 
sense, we obtain the renormalized free energy (with one end fixed at the
origin)
\begin{equation}
\overline{F[0,t;h]} = \overline {F_0(0,t)}-\frac{h^2}{2\kappa}t 
- u \int^t_0 dz \overline{\langle\delta[\xi(z)+h(t-z)/\kappa]\rangle}_{0,t}
\end{equation}
to the lowest order in $u$. Note that 
 $\overline{F_0}$ does not depend on $h$ due to the statistical symmetry.
Remembering that $\overline{ \langle\langle [\xi(t) - \xi(0)]^2 
\rangle\rangle^c_{x,t} }
= - T \partial_h^2 \overline{F}|_{h=0}$ 
where $\langle\langle\ldots\rangle\rangle$
denotes thermal average using the full Hamiltonian ${\cal H}$,  and
defining the renormalized stiffness constant from 
\begin{equation}
\overline{ \langle\langle [\xi(t) - \xi(0)]^2 \rangle\rangle^c_{0,t} }
\equiv \frac{T}{\widetilde{\kappa}} t,
\end{equation}
we have
\begin{eqnarray}
&\widetilde{\kappa}^{-1} &= \kappa ^{-1}- u\frac{\partial^2}
{\partial h^2} \int_0^t \frac{dz}{t} 
G_t^{(1,1)}(-h(t-z)/\kappa,t-z)|_{h=0} \label{rg1.b} \\
& &= \kappa^{-1} ( 1 -  C_\kappa g_0 ). \label{rg1.a}
\end{eqnarray}
In Appendix B, we find $C_\kappa$ to be a finite constant
given by Eq.~(\ref{Ckappa}).
Thus, there is only a {\it finite} renormalization 
of $\kappa$ which could be absorbed in its redefinition.

Next, we consider the renormalization of
 the amplitude of the mean square displacement of the polymer
\begin{equation}
\widetilde{X}^2 \equiv \int dr \ r^2 G_t^{(1,1)}(r,t).
\end{equation} 
Expansion to the lowest order in $u$ yields 
\begin{eqnarray}
&\widetilde{X}^2 \equiv (\widetilde{B} t^\zeta)^2
&=  \int^\infty_{-\infty} dy \ y^2 \left[ G^{(1,1)}_t(y,t) + \frac{1}{T}
\overline{\langle\delta[\xi(0)-y]{\cal H}_1\rangle^c_{x,t}}\right] \nonumber \\
& &=(Bt^\zeta)^2 + u \int^\infty_{-\infty} dy y^2 \int_0^t dz 
G_t^{(2,1)}(0,t-z;y,t) \label{rg3.a}
\end{eqnarray}
where
\begin{equation}
G_t^{(2,1)}(x-y_1,t-z_1;x-y_2,t-z_2) 
= -\frac{1}{T} \ \overline{\langle\delta(\xi(z_1)-y_1)
\delta(\xi(z_2)-y_2)\rangle^c_{x,t}} \label{G21}
\end{equation}
is a nonlinear response function of the noisy-Burgers' equation.
In Appendix A, we obtain an expression for $G^{(2,1)}$ in term of
$G^{(1,1)}$ using the mode coupling approximation, the validity of which
will be discussed shortly.
Using the relations (\ref{response.A}) and (\ref{G21.A}),
 we obtain from Appendix B
\begin{equation}
\widetilde{B} = B(1-C_B g_0),  \label{rg3.b}
\end{equation}
where $C_B$ is another finite constant 
given in Eq.~(\ref{CB}). As for $\kappa$, the small correction to 
$B$ can be absorbed in its definition.

To check the above result in a different way, we calculated also 
 corrections to $\Delta F_0$, or the relation 
(\ref{fdt}), due to the perturbation ${\cal H}_1$. 
Expanding the total free energy  to the lowest order in $u$ gives 
\begin{equation}
\overline{\partial_x F(x,t) \partial_y F(y,t)}
=\frac{\Delta\kappa}{T} \delta(x-y)  
-u\frac{\partial ^2}{\partial x \partial y} \int_0^t dz
G_t^{(1,2)}(x-y,0;\frac{x+y}{2},t) \label{rg2.a}
\end{equation}
where
\begin{equation}
G_t^{(1,2)}(x-y,0;\frac{x+y}{2},t) 
= \overline{F_0(x,t)\langle\delta[\xi(z)]\rangle}_{y,t} 
+\overline{F_0(y,t)\langle\delta[\xi(z)]\rangle}_{x,t} \label{G12} 
\end{equation}
is another three-point response function of 
the noisy-Burgers equation. Again, we use
the mode-coupling approximation to relate $G^{(1,2)}$ to the elementary
functions $G^{(1,1)}$ and $C$ in a simple way. 
Using the relations (\ref{response.A}), (\ref{fdt.A}), and 
(\ref{G12.A}) , we obtain
\begin{equation}
\overline{\partial_x F(x,t) \partial_y F(y,t)} = \frac{\Delta\kappa}{T}
\left[\delta(x-y) - g_0 f(x,y)\right], \label{rg2.b}
\end{equation}
with $f(x,y) \sim |x-y|^{-1}$
if $x + y =0$, and a faster decay for $x+y \ne 0$~\cite{note.1}. Thus, 
the perturbation ${\cal H}_1$
changes the form of the correlation function. However,
 it does not lead to a 
{\it singular} renormalization of $\Delta F_0$, which is obtained by
integrating Eq.~(\ref{rg2.b}) over {\it both}
 the $x$ and $y$ coordinates.

Finally, we consider the renormalization of $u$ itself. This follows
from the expansion of the disorder averaged free energy up to second 
order in $u$: 
\begin{eqnarray}
&\delta F(x,t) &\equiv\overline{F(x,t)}-\overline{F_0(x,t)} \nonumber\\
& &=\overline{\langle{\cal H}_1\rangle}_{x,t}
- \frac{1}{2T}\overline{\langle{\cal H}^2_1\rangle^c_{x,t}}.
\end{eqnarray}
It will be convenient to integrate over the fixed point $x$.
Using the normalization of $G^{(1,1)}$, $\int_{-\infty}^\infty
 dx G^{(1,1)}_t(x,\tau) = 1$, we have
\begin{equation}
\int dx \delta F(x,t) 
= - u t + \frac{u^2}{2}\int_0^t dz_1 dz_2 \int_{-\infty}^\infty dx
G_t^{(2,1)}(x,t-z_1;x,t-z_2). \label{rg4.a}
\end{equation}

Using Eqs.~(\ref{response.A}) and (\ref{G21.A}),  
we get  the following correction to $u$
\begin{equation}
\widetilde{u} = u [ 1  +   C_u g_0 \log (t/a_\|) ]  \label{rg4.b}
\end{equation}
where  $C_u$ is a finite positive constant given by Eq.~(\ref{Cu}), 
and $a_\| \approx (a/B)^{3/2}$ is the short distance cutoff along the length
of the directed polymer (see Appendix B).
The logarithmic divergence in Eq.~(\ref{rg4.b}) indicates the breakdown of 
the small-$u$ perturbation for  $g_0 \log (t/a_\| ) \gg 1$. Thus
the pinning potential is {\it strongly relevant} beyond the localization length
\begin{equation}
 l_{\|}= a_\| e^{1/g_0}, \label{lperp}
\end{equation}  and the directed polymer becomes localized to the line defect.

Since the only quantity that suffers nontrivial renormalization is the
pinning strength $u$, it is straightforward to form a
renormalization-group recursion relation which allows one to
obtain information about the  polymer in dimensions $d > d_c$
 using the calculations performed at the critical dimensionality 
$d_c = 1+1$. 
From Eq.~(\ref{rg4.b}) and with the rescaling  
$ t'= b t$, we obtain the recursion relation 
\begin{equation}
b\frac{dg}{db} = \epsilon(d) g + C_u g^2 \label{rr}
\end{equation}
where $\epsilon(d) = 2 - \zeta(d) (d+1)$ as given in Eq.~(\ref{eps}). 
Thus the coupling constant $g(b)$,
which is a dimensionless measure of  the  pinning strength at scale $b$,
flows to large values from any non-zero initial value, 
leading to the pinning of the directed polymer to the line defect 
at large scales if $\epsilon \ge 0$,
i.e., if $d \le 1+1$. But for $\epsilon < 0$ or $d > 1+1$, 
the pinning potential is
only effective if its strength exceeds a critical value, $g_c = \epsilon/C_u$.
Thus the phase diagram of the directed polymer is in fact given by
Fig.~4(b), like that of the pure problem (Fig.~2),
rather than Fig.~4(a) as suggested by naive scaling arguments
and the uncontrolled Hartree calculation.

\subsection{The Depinning Transition}

We now investigate the critical behavior of the directed polymer close
to the depinning transition. Right at the depinning point $g_c$,
the wandering exponent $\zeta_c$ and the energy exponent $\theta_c$
are simply 
\begin{equation}
\zeta_c = \zeta(d) \qquad {\rm and} \qquad \theta_c =
\theta(d) \label{expc}
\end{equation}
to $O(\epsilon)$ since none of the parameters $\kappa$, $A$ and $B$
pick up divergent renormalization.
The divergence of the correlation lengths, as one approaches the depinning
transition from the pinned side, can be read off from Eq.~(\ref{rr}).
We obtain 
\begin{equation}
l_\| =a_\| \left(\frac{g_0}{g_c}-1\right)^{-\nu_\|} \qquad {\rm and} \qquad
l_\perp =a \left(\frac{g_0}{g_c}-1\right)^{-\nu_\perp}
\end{equation} 
with the liberation exponents
\begin{equation}
\nu_{\|} = 1 / |\epsilon|, \qquad {\rm and} \qquad \nu_\perp = \zeta_c \nu_\| =
\zeta(d)/|\epsilon|, \label{nu}
\end{equation}
again valid to $O(\epsilon)$.
In $d=2+1$ where $\zeta \approx
5/8$ and $\epsilon = 1/2$, 
Eq.~(\ref{nu}) gives $\nu_\perp \approx 1.25$ which is comparable to
the results of numerical simulations: $\nu_\perp =1.3\pm 0.6$ by Balents
and Kardar~\cite{bk1}, and $\nu_\perp =1.8\pm 0.6$ by Tang 
and Lyuksyutov~\cite{tl}. It will be
interesting to see whether Eqs.~(\ref{expc}) and (\ref{nu}) might be valid
 to all orders in
$\epsilon$ (up to some upper critical dimensions), as they do in the case of the pure problem without point disorders~\cite{kolo2}.
This can be directly probed numerically in $d=1+1$ by using correlated point
disorder or by modifying the form of $U_p$ (see Sec.~V).  

The results Eqs.~(\ref{expc}) and (\ref{nu}) have been conjectured earlier 
by Kolomeisky and Staley~\cite{kolo1,kolo2}, 
who combined the renormalization-group flow equations
for the pure depinning problem with $\eta = 0$ and the one for only point
disorder with $U_p=0$ in an {\it ad hoc} way.
In particular, Kolomeisky and Straley neglected to consider 
the renormalization of 
the stiffness $\kappa$, $B$, and the amplitude of the disorder potential $\Delta$ due to the pinning potential $U_p$. It is the lack of any divergent
renormalization of these quantities, as obtained through explicit calculations
in this study,  that ensures
the validity of Eq.~(\ref{expc}), at least to $O(\epsilon)$. 
The lack of such divergent  renormalization follows
from the large transverse wandering of
the polymer at the depinning transition, thus diminishing the effect
of the pinning potential. (Similar behaviors have been found 
for the renormalization of the surface tension of a wetting layer at the 
wetting transition~\cite{wetting}.)

Balents and Kardar~\cite{bk2} obtained  similar conclusions, Eqs.~(\ref{expc})
and (\ref{nu}), by generalizing the directed polymer to a $D$-dimensional
manifold, and then computing the analogy of the coefficients $C_\kappa$,
$C_B$ and $C_u$ in a functional renormalization group (FRG) analysis to first
order in $\delta = 4- D$. Since the $O(\epsilon)$ results  (\ref{expc})
and (\ref{nu}) do not depend on the numerical values of the $C$'s, the FRG
approach is  effective   despite the large expansion
parameter ($\delta = 3$) for the directed polymer.

Finally, we comment on the validity of the mode coupling approximation
used in the evaluation of the expressions leading to Eqs.~(\ref{rg1.b}),
(\ref{rg3.b}), (\ref{rg2.b}) and (\ref{rg4.b}). Due to the combination of
a statistical tilt symmetry and a fluctuation dissipation theorem in
$d=1+1$, the mode coupling approach gives the correct {\it scaling}
 behavior for
the functions $G^{(m,n)}$ (see Refs.~\cite{mode,mode2} and Appendix A). 
This is all that is needed here since, as explained above, the results
(\ref{expc}) and (\ref{nu}) are independent of the numerical values of the
coefficients $C$'s. On the other hand, the mode coupling method is known
to give very good quantitative results even for the scaling
functions~\cite{hhk,mode,mode2}. Thus
the coefficient $C$'s computed in this way should be quantitatively accurate.
 
[ Note added: Very recently, Kolomeisky and Staley~\cite{kolo3} modified their
renormalization-group analysis and obtained a power-law rather than exponential 
divergence of the localization length at the critical dimension $1+1$. 
We disagree with their conclusions and point
out what we believe to be the key difference between Ref.~\cite{kolo3} and the 
present work: The  
analysis of Ref.~\cite{kolo3} is a one-loop calculation with respect to
the {\it pure} directed polymer. 
The effect of point disorder is taken into account
by an {\it Ansatz} which makes appropriate rescaling of the coefficients.
On the other hand, the analysis presented
in this section {\it starts} from the low-temperature fixed point of
 the randomly-pinned directed polymer, and takes into account of the extended
pinning potential in a {\it systematic} treatment. 
The two approaches should eventually lead to the same scaling behavior
 (see Sec.~IV.D).
However, the approach of Ref.~\cite{kolo3} requires a careful, consistent
formulation. In particular, the choice of parameters [Eq.~(2.4)] used in Ref.~\cite{kolo3} implies that not all ``temperatures'' renormalize in the
same way (since Galilean invariance must be respected). This introduces some
ambiguities in the scaling ansatz used there. ]

\subsection{Physical Interpretations}

As shown by the renormalization group analysis, the logarithmic divergence
of the effective pinning strength $\widetilde{u}$ at the critical dimension 
is the single most important element leading to the phase diagram 
Fig.~4(b) and the exponents (\ref{expc}) and (\ref{nu}). Discussions 
of the preceding paragraphs also illustrates the robustness of these results
--- the same results are obtained from a variety of methods, some with 
drastic  unjustified approximations, as long as the {\it scaling} behaviors 
are properly included. We shall now provide a physical picture of the weak
localization using a phenomenological scaling theory, 
and argue that the logarithmic divergence of $\widetilde{u}$ should indeed 
be expected at the critical dimension.

In Sec.~II, we established the energy gained by the directed polymer
(due to the presence of a  weak pinning potential) to be
 $\delta F_0 \propto u \int_0^t d\tau (B\tau)^{-d_\perp\zeta}\sim t^{1/3}$
in $1+1$ dimensions. This corresponds to describing the optimal path
of the directed polymer as a generalized random walk (with the wandering
exponent $\zeta$ rather than $1/2$), and then equating the energy gained to
the accumulated return probability of the random walk. However, such a
treatment of the optimal path is oversimplified. As discussed in detail in
Ref.~\cite{hf}, while the optimal path in a typical sample (or a typical
region of a very large sample) is unique, there is a nonzero probability
that there exists a different path whose free energy is within $O(1)$
of that of the optimal path. The probability of finding two such paths
a distance $\Delta$ apart is $p(\Delta) \sim \Delta^{-3/2}$ in $1+1$
dimension~\cite{hf}, 
and the ``droplet'' formed (i.e., the difference between 
the two paths) typically have a length $\tau \sim \Delta^{1/\zeta}$,
which is of $O(\Delta^{3/2})$ in $1+1$ dimension.  Thus, if one of the
optimal path encounters the origin, then the probability of having a nearly
degenerate optimal path also encountering the origin is $p(\tau^{2/3}) \sim
\tau^{-1}$ in $1+1$ dimensions.
In this way, we see that the {\it accumulated} effect of the
statistically droplets existing at different scales 
leads to a logarithmic divergence, i.e., $\int^t d\tau p(\tau^{2/3}) \sim
\log t$. This makes the columnar pin
marginally relevant at the critical dimension, and results in a phase diagram
of the type depicted in Fig.~4(b) rather than that in Fig.~4(a). Note that the
important ingredients  of the above argument are only the probability
of droplet formation and the shape of the droplets.
It therefore suggests the marginal relevance of an extended defect 
at the critical dimension to be
 a general consequence of the droplet scaling theory.

Technically, the droplets manifest themselves in the expression describing the
renormalization of the pinning strength in Eq.~(\ref{rg4.a}), a diagrammatic
representation of which is shown in Fig.~6. The similarity of the droplet
configuration in Fig.~5 and the loop diagram in Fig.~6 is striking.
 As explained in detail in Ref.~\cite{hf}, the distribution of droplets
is given by the  function $G^{(2,1)}$. However, the particular distribution
discussed in Ref.~\cite{hf} has to do with droplets with a given width
$\Delta$ at the same vertical coordinate, whereas the distribution 
needed here is the one with a given length $\tau$
 at the same transverse coordinate. Arguments leading to the logarithmic
divergence in the preceding paragraphs
assumes $\tau \sim \Delta^{1/\zeta}$. This is validated by the more detailed
calculation given in Appendix B.

In some ways, the depinning transition here is similar to the 
de Almeida-Thouless~\cite{AT} line in a spin glass. The columnar pin which
breaks the statistical translational symmetry of the directed polymer
plays a role similar to the external magnetic field which breaks the
statistical up-down symmetry of the Ising spin glass. It should be
interesting to study the depinning transition in the replica formalism.
The encounter of replica symmetry breaking alluded to in Sec.~III
 is no longer mysterious now --- it is simply how the droplet excitations
are manifested within the replica formalism~\cite{mp,hf}.

\vfill
\newpage 

\section{Related  Depinning Problems}

The results of Sec.~IV can be easily generalized to long range correlated 
point disorders, other forms of pinning potentials, 
and higher dimensional elastic manifolds. Here we shall describe a few
interesting cases to illustrate the method.

Following  Nattermann~\cite{nat2} and Medina {\it et al}~\cite{medina},
 we consider a random
potential with long-range correlations of the form
\begin{equation}
\overline{\eta (\bx,z) \eta (\bx', z')} =  2 \Delta \delta (z - z')
{|\bx -\bx'| }^{2\rho - d +1}. \label{corr}
\end{equation}
We shall be focus on the case $d=1+1$, where the model (\ref{H0}) 
 describes the domain wall of the low temperature phase of
 a random Ising model in two dimensions. 
The correlator (\ref{corr}) extrapolates
smoothly between the case of random bond ($\rho=0$) and random field
($\rho=1$)~\cite{medina,nat2}.
It is known that the roughness exponent $\zeta (\rho)$ stays 
at its value of $2/3$
 for short range correlated disorder as long as $\rho \le 1/4$. 
For $ 1 \ge \rho \ge 1/4$ it takes on the Flory 
value of $\zeta(\rho) = 3/{(5-2\rho)}$. From Eq.~(\ref{rr}) and $\epsilon
= 2-\zeta(d+1)$, we see that $\epsilon = 0$ for $\rho \le 1/4$, and
$\epsilon = (1-4\rho)/(5-2\rho) < 0$ for $\rho > 1/4$. Thus for $\rho >
1/4$, a  small pinning 
potential is irrelevant. As one varies $u$, 
there will be a depinning transition with  
exponents $ \nu_\| = 1/|\epsilon| = (5-2\rho)/(4\rho-1)$ and
$\nu _{\perp} = \nu _{\|}\zeta_c = 3/{(4 \rho - 1)}$ to
$O(\epsilon)$. If the exponent $\nu_\|=1/|\epsilon|$ is indeed exact to
all orders as in the pure case, then the above expression will be valid
for all $1\ge \rho > 1/4$.
 In particular we should obtain a depinning
transition  with the 
exponents $\nu _{\perp} = \nu _{\|} = 1$ for $\rho=1$.
A numerical simulation of the depinning transition  
of the domain wall of the 2d random field Ising model should therefore
be an efficient way of probing the ``exactness" of the $O(\epsilon)$ result.

We can generalize the methods described in this paper to study the
pinning of a directed polymer by other forms of the pinning potential
$U_p(\br)$.  If we write ${\cal H}_1 = \int d^{d_\perp}\br U_p(\br) 
\int_0^t dz \delta^{d_\perp}[\br -\bxi(z)]$,
then the lowest order correction to the free energy is 
\begin{equation}
\delta F_0(t)  = \overline{\langle{\cal H}_1 \rangle}
= \int_0^t dz \int d^{d_\perp}\br \ U_p(\br)\ G^{(1,1)}_t(\br,t-z), \label{df2}
\end{equation}
where $G^{(1,1)}$ is the one-point distribution function defined in
Eq.~(\ref{G}), with the scaling properties described in Appendix A.
Suppose $U_p(\br)$ has a long tail, i.e., $U_p = u/r^s$,
then $\delta F_0(t) \sim t^{1-s\zeta}$. Comparing this to $\Delta F_0 \sim
t^{\theta}$ and recalling the exponent identity Eq.~(\ref{id}), 
we find the critical dimension of $U_p$ to be
\begin{equation}
2 = \zeta(d_c) (2+s).
\end{equation} We expect a depinning transition at finite $u$ above 
the critical dimension $d_c$, with the liberation exponent $\nu_\| =
1/|2-\zeta(d)(2+s)|$.

We may also consider a pinning potential $U_p(\br) = -u \delta^{d_\perp}
[\br - {\bf R}(z)]$, where the trajectory of the pin ${\bf R}(z)$ is
itself an arbitrary (but fixed) function of $z$. For example, we may
have $R(z) \sim z^{\zeta_R}$, describing the quenched defect 
 trajectory of a superconductor subject to random collision by heavy
ions.   In this case,
Eq.~(\ref{df2}) becomes 
\begin{equation}
\delta F_0(t) = - u \int_0^t dz \ G_t({\bf R}(z),t-z).
\end{equation}
If $R(z) < |t-z|^\zeta$ or $\zeta_R < \zeta$, 
i.e., if the transverse fluctuation of the defect trajectory is smaller
than that of the randomly pinned directed polymer,
 then the $z$-dependence of the pinning potential
is irrelevant and we have $\delta F_0(t) \sim u t^{1- d_\perp \zeta}$
as before, and the defect acts like a straight columnar pin.
 However, if $R(z) > |t-z|^\zeta$, as for example is the case
for a misoriented columnar pin where $R(z) \propto |t-z|$~\cite{splay}, 
then $G^{(1,1)}$ is
sharply cutoff and $\delta F_0(t)$ becomes finite for large $t$. In this
case, a weak pinning potential is irrelevant compared to the random energy
gain $\Delta F_0 \sim t^\theta$.
This type of analysis can be extended to 
study a large variety of pinning potentials, including the case where
${\bf R}(z)$ itself is the trajectory of another directed
polymer~\cite{hwa}.

Another interesting ramification of our considerations 
pertains to $D$-dimensional oriented
manifolds in $d$ embedding dimensions, i.e., $\vec{z} \in \Re^D$ and
$\bxi(\vec{z}) \in \Re^{d-D}$. $D=1$ describes a directed polymer or a flux
line, and $D=2$ describes, for example, the domain wall of a 
three dimensional ferromagnet. The free 
energy scales in this case  with an exponent $\theta = 2\zeta+ D-2$,
which is the $D$-dimensional analog of the exponent identity (\ref{id}).
Following Ref.~\cite{bk1}, we also generalize the extended defect to be 
$n$-dimensional, i.e.,
$U_p(\br) = -u\delta^{d-n}(\br)$, and ${\cal H}_1 = \int d^D \vec{z} \ 
U_p[\bxi(\vec{z})]$. The columnar pinning potential discussed corresponds
to $n=1$, and planar defects such as grain boundaries correspond to $n=2$.
  Straightforward generalization of the scaling
arguments of Sec.~II and Eq.~(\ref{dF}) 
gives $\delta F_0(t) = \overline{\langle{\cal H}_1 \rangle} \sim - u t^{D
- \zeta(d-n)}$, where $t$ is now the linear size of the manifold.
Comparing this to $\Delta F_0 \sim t^\theta$, we find the 
condition for the 
relevance of $U_p$ to be given by 
\begin{equation}
2 > \zeta (d)(2 + d - n). \label{dc.2}
\end{equation}
An important application here is the pinning of an interface to a planar
defect, say the localization of a domain wall to a grain boundary in a 
random ferromagnet. In this case, $D=n = 2$, $d=3$, and Eq.~(\ref{dc.2})
 reads $\zeta (3) < 2/3$. 
For the interface of a random bond Ising magnet,
 this is always fulfilled and the interface is 
always pinned by the planar defect. For random field systems however, 
$\zeta(3) = 2/3$ exactly in 3 dimensions,
 and our criterion (\ref{dc.2}) is inconclusive.
Straightforward generalization of the scaling argument given in Sec.~IV.D
shows that the interface is again weakly
pinned by an arbitrary weak pinning potential.

\newpage
\acknowledgements
We are grateful to many helpful discussions with L. Balents, D. S. Fisher,
T. Halpin-Healy, M. Kardar, E. B. Kolomeisky, J. Krug, D. R. Nelson, 
and   L.-H. Tang. 
TH is supported  by US Department of Energy Contract No.~DE-FG02-90ER40542.
TN acknowledges the financial support of the Volkswagen Foundation and
the hospitality of Harvard University where a part of this work was done.

\vfill

\newpage
\appendix
\section{Useful Relations for the directed polymer}

In order to make the paper more self-contained,  we summarize  
here some useful 
results on the randomly pinned directed polymer
 obtained in earlier publications~\cite{hf,medina,mode}.
The restricted partition function of a directed polymer
 of length $t$ with one end  $\bxi(t)$ fixed at $\bx$ is 
\begin{equation}
Z_0(\bx,t) = \int {\cal D}[\bxi]\delta^{d_\perp}[\bx-\bxi(t)]
e^{-{\cal H}_0/T}.
\end{equation}
The free energy $F_0(\bx,t) = -T \log Z_0(\bx,t)$
fulfills the noisy-Burgers' equation~\cite{fns,kpz,medina}
\begin{equation}
\partial_t F_0(\bx,t) = \frac {T}{2\kappa}{\grad}^2 F_0
- \frac{1}{2\kappa}(\grad F_0)^2 + \eta(\bx,t). \label{kpz.A}
\end{equation}

In Sec.~IV, we introduced various  correlation and distribution
 functions to evaluate
the effect of the pinning potential ${\cal H}_1$. To begin with, the 
one-point distribution function is 
\begin{equation}
\overline{\langle\delta[\bxi (z) - \by]\rangle}_{\bx,t}
= G_t^{(1,1)} ( \bx - \by , t-z ).
\end{equation}
Again, $\langle\ldots\rangle_{\bx,t}$ denotes thermal average using
the partition function $Z_0(\bx,t)$ and the subscript $t$ is used
in $G^{(1,1)}$ to denote the explicit $t$-dependence of the distribution
function. By simple scaling and normalization requirements, we have
\begin{equation}
G^{(1,1)}_t(\br,\tau) = (B\tau^\zeta)^{-d_\perp}
\tg_{t/\tau}(r/B\tau^\zeta),\label{response.A}
\end{equation}
where $\tg_{t/\tau}(0)$ is finite, $\tg_{t/\tau}(s)$ decreases sharply 
for  $s \gg 1$, and $\int d^{d_\perp} s \ \tg_{t/\tau}(s) = 1$
 for all $t/\tau$.
The rms displacement $X^2_t(\tau)$ is given by the second moment,
\begin{eqnarray}
&X^2_t(\tau) &\equiv \int d^{d_\perp}\br\ \br^2 G_t^{(1,1)}(\br,\tau)
\nonumber\\
&  &= B^2 \tau^{2\zeta} \int d^{d_\perp}s\ s^2 \tg_{t/\tau}(s),
\end{eqnarray}
which is a weak function of $t/\tau$. It is found
numerically~\cite{kim,hhk}
 that $X^2_t(\tau)/\tau^{2\zeta}$ is the same order of magnitude 
for $t = \tau$ and $t \gg \tau$.
We fix $B$ by the rms displacement of the free end, $X^2_t(t) = B^2
t^{2\zeta}$ (see Eq.~(\ref{Xrms})). This is accomplished by choosing
\begin{equation}
\int d^{d_\perp}s \ s^2 \tg_1(s) = 1. \label{width.A}
\end{equation}

Other than $G^{(1,1)}$, we are also interested in the higher-order
distribution functions,
\begin{eqnarray}
&G_t^{(2,1)}(\bx-\by_1,t-z_1;&\bx-\by_2,t-z_2) = -\frac{1}{T} 
\overline{\langle\delta[\bxi (z_1) - \by_1] \delta[\bxi (z_2) - \by_2 ]
\rangle_{\bx,t}^{(c)}}\ ,\\
&G_{[t_1,t_2]}^{(1,2)}(\bx_1-\bx_2,t_1-t_2;&\frac{\bx_1+\bx_2}{2}-\by,
\frac{t_1+t_2}{2}-z) \nonumber \\
& &= \left[\overline{F_0(x_1,t_1)\langle\delta[\xi(z)-\by] 
\rangle_{\bx_2,t_2}} 
+ \overline{F_0(x_2,t_2)\langle\delta[\xi(z)-\by]
\rangle_{\bx_1,t_1} } \right] ,
\end{eqnarray}
as well as the free energy correlation function~\cite{corr.note}
\begin{equation}
C_{[t_1,t_2]}(\bx_1-\bx_2,t_1-t_2) 
= \overline{ [F_0(\bx_1,t_1) - F_0(\bx_2,t_2)]^2 },
\end{equation}
where the subscript $[t_1,t_2]$ denotes the larger of $t_1$ and $t_2$.
In analogy to $G^{(1,1)}$, the correlation function has a similar scaling
form,
\begin{equation}
C_t(\br,\tau) = A^2 t^{2\theta} \tc_{t/\tau}(r/B\tau^\zeta).
\end{equation}
The scaling function $\tc$ gives
the following scaling properties for $C$,
\begin{equation}
C_t(\br,\tau)  \left\{ 
\begin{array}{ll}
= 2 A^2 t^{2\theta} & B\tau^\zeta  \ll B t^\zeta \ll r, \\
\propto 2 A^2 (r/B)^{2\theta/\zeta} & B\tau^\zeta \ll r \ll B t^\zeta, \\
\propto 2 A^2 \tau^{2\theta} & r \ll B\tau^\zeta \ll B t^\zeta.
\end{array}
\right. \label{corr.A}
\end{equation}
For convenience, we also write the correlation function as
\begin{equation}
C_{[t_1,t_2]}(\bx_1-\bx_2,t_1-t_2) = G^{(0,2)}_{t_1}(0,0) +
G^{(0,2)}_{t_2}(0,0) - 2 G^{(0,2)}_{[t_1,t_2]}(\bx_1-\bx_2,t_1-t_2) \label{G02}
\end{equation}
where
\begin{equation}
G^{(0,2)}_{[t_1,t_2]}(\bx_1-\bx_2,t_1-t_2) 
= \overline{F_0(\bx_1,t_1)F_0(\bx_2,t_2)}.
\end{equation}
The scaling properties of $G^{(0,2)}_t(\br,\tau)$ is easily obtained from Eq.~(\ref{corr.A}). For example, from Eq.~(\ref{G02}) and $G^{(0,2)}_t(\br\to \infty,\tau) \to 0$, we have
$G^{(0,2)}_t(0,0) = \frac{1}{2}C_t(\infty) = A^2 t^{2\theta}$. 
Below, we shall provide the approximate forms for all of the $G^{(m,n)}$'s.

It was shown in Ref.~\cite{hf} that $G^{(m,n)}$ can be
obtained simply by adding a source term $\tJ(\bx,t)$ to 
right hand side of the equation of
motion (\ref{kpz.A}) and then taking appropriate derivatives,
i.e., 
\begin{equation}
G^{(m,n)} = \overline{\frac{\delta}{\delta\tJ(\by_1,z_1)}\ldots
\frac{\delta}{\delta\tJ(\by_m,z_m)} [F_0(\bx_1,t_1) \ldots F_0(\bx_n,t_n)]}.
\end{equation}
In the context of the stochastic hydrodynamics of the noisy-Burgers'
equation, the above is nothing but the generalized 
{\it response function}.  In Ref.~\cite{hf}, it was shown that the
nonlinear response function $G^{(2,1)}$ gives the statistics of the rare but
singular ``droplet excitations'', which are connected to replica symmetry
breaking in the replica formalism~\cite{p,mezard}. Here we encounter
them again in perturbation theory, as we already did when using 
the replica formalism in Sec.~III.

However, unlike Sec.~III where we failed to develop a perturbative
expansion due to the lack of knowledge of replica-symmetry broken excited
states, here we can construct the forms of the nonlinear response functions
$G^{(m,n)}$ rather straightforwardly by exploiting a Fluctuation
Dissipation Theorem (FDT), which the equation of motion (\ref{kpz.A}) 
satisfies in $1+1$ dimensions~\cite{medina,mode}.
For example,  the FDT gives
\begin{equation}
\frac{\partial}{\partial x} \frac{\partial}{\partial y}
G^{(0,2)}_t(x-y,\tau)  
= \frac{\Delta\kappa}{T} G^{(1,1)}_t(x-y,\tau), \qquad \tau > 0. \label{fdt.A}
\end{equation}
Taking the limit $\tau \to 0$ in Eq.~(\ref{fdt.A}) and using the
definition of $G^{(1,1)}$, we immediately obtain
\begin{equation}
\overline{\partial_x F_0(x,t) \partial_y F_0(y,t)} = \frac{\Delta\kappa}{T} \delta(x-y).
\end{equation}
Similarly, by integrating Eq.~(\ref{fdt.A}) and using the scaling forms
for $G^{(1,1)}$, we can recover the scaling properties of $C$ given in
Eq.~(\ref{corr.A}).

The combination of the FDT and a
Galilean invariance (corresponding to the statistical rotational symmetry)
in $1+1$ dimensions allows one to use 
a mode-coupling scheme~\cite{mode,mode2,bks} to obtain the forms of the
functions $G^{(m,n)}$. In particular,
 $G^{(1,1)}$ and $G^{(0,2)}$ are given, to a very good
approximation, by the following set of self-consistent integral equations,
\begin{eqnarray}
&G_t^{(1,1)}(x-y,t-z) &= \hG^{(1,1)}(x-y,t-z)  \nonumber\\
& &+ \frac{1}{\kappa^2} \int^\infty_{-\infty} dx' dy' 
\int^t_0 dt' \int_0^{t'} dz'\hG^{(1,1)}(x-x',t-t')
\frac{\partial}{\partial x'}G_{t'}^{(1,1)}(x'-y',t'-z') \nonumber \\
& &\quad\frac{\partial}{\partial x'}\frac{\partial}{\partial y'} 
G_{t'}^{(0,2)}(x'-y',t'-z') 
\frac{\partial}{\partial y'} G_{z'}^{(1,1)}(y'-y,z'-z), \label{mc.1}\\
&G_{[t_1,t_2]}^{(0,2)}(x_1-x_2,t_1-t_2) &=
\hG^{(0,2)}_{[t_1,t_2]}(x_1-x_2,t_1-t_2) \nonumber \\
& &+ \frac{1}{2\kappa^2} \int_{-\infty}^\infty dx_1' dx_2' \int^{t_1}_0 dt_1' 
\int_0^{t_2} dt_2' G^{(1,1)}_{t_1}(x_1-x_1',t_1-t_1') \nonumber \\
& &\quad G^{(1,1)}_{t_2}(x_2-x_2',t_2-t_2')
\left[\frac{\partial}{\partial x_1'}\frac{\partial}{\partial x_2'} 
G_{[t_1',t_2']}^{(0,2)}(x_1'-x_2',t_1'-t_2') \right]^2, \label{mc.2}
\end{eqnarray}
where 
\begin{equation}
\hG^{(1,1)}(r,\tau) = \sqrt{\frac{T}{2\pi\kappa\tau}} 
\exp\left[-\frac{T}{\kappa}\frac{r^2}{\tau}\right], \qquad \tau > 0
\end{equation}
is the ``bare'' response function, and
\begin{equation}
\hG^{(0,2)}_{[t_1,t_2]}(x_1-x_2,t_1-t_2) 
= 2\Delta \int_{-\infty}^{\infty} dx' \int_0^{t_2}dt' 
\hG^{(1,1)}(x_1-x',t_1-t')\hG^{(1,1)}(x_2-x',t_2-t')
\end{equation}
is the ``bare" correlation function.  The mode-coupling
equations (\ref{mc.1}) and (\ref{mc.2}) can be solved by using the
scaling forms (\ref{response.A}) and (\ref{corr.A}) for $G^{(1,1)}$ and $G^{(0,2)}$.
The scaling functions $\tg$ and $\tc$ obtained in this way
are in very good agreement with those from numerical
simulations~\cite{kim,hhk,mode}.
It is found that $\tg_\sigma(s)$  is  approximately a Gaussian with
a weak $\sigma$ dependence, and the width of the ``Gaussian" is fixed
by the condition (\ref{width.A}) to be $1$.

The functions $G^{(1,1)}$ and $G^{(0,2)}$ can now be used to construct
higher order response functions via the mode-coupling scheme. For instance,
\begin{eqnarray}
&G_t^{(2,1)}(x-y_1,t-z_1;&x-y_2,t-z_2) \nonumber \\
& &=-\frac{1}{\kappa}\int_{-\infty}^\infty dx' \int_0^t dt' 
G_t^{(1,1)}(x-x',t-t') \nonumber \\
& & \quad \frac{\partial}{\partial x'}G_{t'}^{(1,1)}(x'-y_1,t'-z_1)
\frac{\partial}{\partial x'}G_{t'}^{(1,1)}(x'-y_2,t'-z_2) \label{G21.A}\\
&G_{[t_1,t_2]}^{(1,2)}(x_1-x_2,t_1-t_2;
&\frac{x_1+x_2}{2}-y,\frac{t_1+t_2}{2}-z) \nonumber \\
& &= -\frac{1}{\kappa} \int_{-\infty}^\infty dx_1' \int_0^{t_1} dt_1'
G_t^{(1,1)}(x_1-x_1',t_1-t_1') \nonumber \\
& & \quad \quad\frac{\partial}{\partial x_1'}G_{t_1'}^{(1,1)}(x_1'-y,t_1'-z)
\frac{\partial}{\partial x_1'}G_{[t_1',t_2]}^{(0,2)}(x_2-x_1',t_2-t_1') 
\nonumber \\
& & \quad + {\rm permutation of} [ (x_1,t_1) \leftrightarrow (x_2,t_2) ]. \label{G12.A}
\end{eqnarray}
The validity of the above  mode coupling approximation for $G^{(2,1)}$
was discussed in detail in Ref.~\cite{hf}. Eqs.~(\ref{G21.A}) and
(\ref{G12.A}) should capture the key scaling properties but may not be
quantitatively accurate. We shall nevertheless use the above expressions
to evaluate the integrals obtained in the RG analysis of Sec.~IV. 
As our main concern is the existence 
of logarithmic divergence in the
renormalization of various parameters, rather than the numerical 
value of any particular integrals, the use of the mode coupling
approximation should be adequate.

\section{Results of Calculations}

In this appendix, we compute the perturbative effect of ${\cal H}_1$
on the parameters $\kappa$, $B$, and $u$, by using the expression for
$G^{(m,n)}$'s obtained from the mode coupling approximation (see Appendix
A). We are particularly interested in the $t$-dependence of the 
renormalized parameters $\tilde\kappa(t)$, $\tilde B(t)$ and $\tilde u(t)$
in the limit $t\to \infty$.

We start with the renormalization of the stiffness $\kappa$. From Eq.~(\ref{rg1.a})
and the scaling form (\ref{response.A}) for $G^{(1,1)}$, we have
\begin{eqnarray}
&\widetilde{\kappa}^{-1} & = \kappa^{-1} - u \int_0^t \frac{dz}{t} 
\frac{(t-z)^2}{\kappa^2[B(t-z)^\zeta]^3} \frac{\partial^2}{\partial s^2}
\tg_{t/(t-z)}(s)|_{s=0} \nonumber \\
& & = \frac{1}{\kappa} -\frac{u}{\kappa^2 B^3} \int_0^1 d\sigma
\tg''_{1/\sigma}(0), \label{tkappa}
\end{eqnarray}
where `primes' indicate derivatives of $\tg$ 
and $\zeta = 2/3$ in $d=1+1$.
Thus we obtain Eq.~(\ref{rg1.b}) with $g_0 = u/(\kappa B^3)$ and 
\begin{equation}
C_\kappa =\int_0^1 d\sigma \tg''_{1/\sigma}(0) \label{Ckappa}
\end{equation}
which is finite.

Next we consider the renormalization of the transverse wandering
coefficient $B$. From Eq.~(\ref{rg3.a}), we have
\begin{equation}
\widetilde{B}^2 t^{2\zeta} = B^2 t^{2\zeta} + I_B,
\end{equation}
with
\begin{eqnarray}
&I_B &=  u \int_{-\infty}^{\infty} dy \ y^2 \int_0^t dz \ 
G_t^{(2,1)}(0,t-z;y,t) \nonumber \\
& &= -\frac{u}{\kappa} \int_0^t dz \int_z^t dt' \int_{-\infty}^\infty dx'
\int_{-\infty}^\infty dy \ y^2 G_t^{(1,1)}(x',t-t') \nonumber \\
& & \quad \frac{\partial}{\partial x'} G_{t'}^{(1,1)}(x'-y,t')
\frac{\partial}{\partial x'} G_{t'}^{(1,1)}(x',t'-z), \label{IB}
\end{eqnarray}
where we used the mode-coupling approximation Eq.~(\ref{G21.A}) for $G^{(2,1)}$.
Using the scaling form (\ref{response.A}) for $G^{(1,1)}$, and noting that $\tg(s)$ is symmetric in $s$, we find
\begin{equation}
I_B = - (B^2 t^{2\zeta}) 2 g_0 C_B
\end{equation}
where
\begin{equation}
C_B =  \int_0^1 \frac{d\tau}{\tau^\zeta} \int_0^1 d\sigma \int_{-\infty}
^\infty ds (-s) \tg'_{(1-\tau)/\sigma}(s) \ 
\tg_{1/\tau}( s \sigma^\zeta/\tau^\zeta). \label{CB}
\end{equation}
Again $C_B$ is finite since $\tg$ are normalized and sharply cutoff for
large argument. We thus obtain the result Eq.~(\ref{rg3.b}).

Finally, we consider the renormalization of $u$. The expression given by Eq.~(\ref{rg4.a}) can be described diagrammatically as in Fig.~6. If we
use the mode-coupling approximation Eq.~(\ref{G21.A}) for $G^{(2,1)}$,
and note the normalization condition  $\int dx G_t^{(1,1)}(x,\tau) = 1$, 
we find
\begin{eqnarray}
& \delta F(t) &= - u t+ \frac{u^2}{2} \int_0^t dz_1 dz_2 
\int_{-\infty}^\infty dx G_t^{(2,1)}(x,t-z_1;x,t-z_2)\nonumber \\
& &= - u t - \frac{u^2}{2\kappa } \int_0^t dt' \int_0^{t'} dz_1 \int_0^{t'} dz_2
 \int_{-\infty}^\infty dx' \frac{\partial}{\partial x'}
G_t^{(1,1)}(x',t'-z_1) \frac{\partial}{\partial x'}
G_t^{(1,1)}(x',t'-z_2).
\end{eqnarray}
Using the scaling form (\ref{response.A}) for $G^{(1,1)}$ again, we obtain
\begin{equation}
\delta F(t) = - u t - \frac{u^2}{2\kappa B^3} \int_0^t dt' I_u(t'),
\label{dFt}
\end{equation}
where
\begin{equation}
I_u(\tau) = \int_0^\tau \frac{dt_1}{t_1^\zeta} \frac{dt_2}{t_2^{2\zeta}}
\int_{-\infty}^\infty ds \ \tg'_{\tau/t_1}(s)
\ \tg'_{\tau/t_2}(s t_1^\zeta/t_2^\zeta).
\label{Iu}
\end{equation}
We shall see that $I_u$ is actually divergent. To regularize the integral,
we insert a ultra-violet cutoff scale $a_\| \propto (a/B)^{1/\zeta}$,
since in our model, the columnar pin $U_p$ is really a potential well of finite size $a$; it is only approximated by a delta function at scales much larger than $a$. Eq.~(\ref{Iu}) then becomes  
\begin{equation}
I_u(\tau) = \int_{a_\|}^\tau\frac{dt_1}{t_1} F(t_1/\tau,a_\|/\tau),
\label{Iu.2}
\end{equation}
where
\begin{equation}
F(\hat{t},\hat{a}) = \int_{\hat{a}/\hat{t}}^{1/\hat{t}} d\sigma
\sigma^{-2\zeta} \int_{-\infty}^{\infty} ds \ \tg'_{1/\hat{t}}(s) 
\ \tg'_{1/(\sigma\hat{t})}(s/\sigma^\zeta).
\end{equation}
Clearly the divergent part of $I_u(\tau)$ comes from the limit
$\tau/a_\|\to 0$ in Eq.~(\ref{Iu.2}). So to leading order,
we have $I_u(\tau) = \log(\tau/a_\|) C_u$, where
\begin{eqnarray}
&C_u &= \lim_{\hat{a}\to 0} F(\hat{a},\hat{a}) \nonumber \\
& &= \lim_{\hat{a}\to 0} \int_1^{1/\hat{a}} \frac{d\sigma}{\sigma^{2\zeta}}
\int_{-\infty}^\infty ds \ \tg'_{1/\hat{a}}(s) 
\ \tg'_{1/(\sigma\hat{a})}(s/\sigma^\zeta) \nonumber \\
& &\approx \int_1^{\infty} \frac{d\sigma}{\sigma^{2\zeta}}
\int_{-\infty}^\infty ds \ \tg'_{\infty}(s) \ \tg'_{\infty}(s/\sigma^\zeta),
\label{Cu}
\end{eqnarray}
which is positive definite since $2\zeta = 4/3 > 1$ and $\tg$ is well behaved.
The effective pinning potential $\tilde u$ can now be defined as
\begin{equation}
\tilde u \equiv -\frac{\partial}{\partial t} \delta F(t) 
= u [ 1 +  g_0 C_u \log(t/a_\|) ],
\end{equation}
which is quoted in Eq.~(\ref{rg4.b}).

\begin{figure}
\caption{ (a) The solid, dashed, and dotted paths are the {\it optimal} 
paths of a randomly pinned directed polymer with one end fixed at $x_1$,
$x_2$, and $x_3$ respectively. The rms displacement of the free end is $X(t)
\propto t^\zeta$, where $t$ is the polymer length. For small separation 
of the fixed end, $x_2-x_1 \ll X(t)$, the two optimal paths typically merge
at a length $\tau \propto |x_2-x_1|^{1/\zeta}$ from the fixed end.
For much larger separations, $x_3-x_1 \gg X(t)$, the optimal paths become
completely uncorrelated.  (b) Variations in the directed polymer's free
energy as a function of the fixed end position, $x$. For large separation
of the fixed end, typical free energy differences are $\Delta F_0 \propto
t^\theta$. 
}
\end{figure}

\begin{figure}
\caption{ The phase diagram of a pure directed polymer pinned 
by a short-ranged columnar pinning well $U_p$. 
The arrows indicate the effective
pinning strength $U$ in units of the entropy cost of localization, 
$U^{(0)}_0=T^2/(2\kappa a^2)$.
The directed polymer is pinned in $d_\perp \le 2$. In $d_\perp > 2$, 
a depinning transition at $U = U^{(0)}_c \propto (d_\perp - 2) U^{(0)}_0$
(indicated by the asterisk) separates the pinned
and depinned phases. }
\end{figure}

\begin{figure}
\caption{ The solid line depicts a directed polymer localized to a columnar
pin at the origin.  The localization distance is $l_\perp$. The polymer is
consisted of a number of uncorrelated segments of length $l_\| \propto
l_\perp^{1/\zeta}$. An optimal path far away from the origin is depicted by
the dashed line.
}
\end{figure}

\begin{figure}
\caption{ (a)  A possible phase diagram for the directed polymer pinned 
by both the columnar pin and a random background. 
The arrows indicate the effective
strength $U$ of the columnar pin, in units of the random energy loss, 
$U_0=A(a/B)^{(\theta-1)/\zeta}$. The critical dimension is $d_\perp = 1$.
Unlike Fig.~2, this phase diagram  contains a depinning transition
in {\it all} dimensions. (b) An alternative phase diagram having the 
same structure as that of the pure one (Fig.~2), but with a shift in the
 critical dimension to $d_\perp = 1$.}
\end{figure}

\begin{figure}
\caption{ The solid line is an optimal path which intersects the origin
at $z=z_1$. The dashed line is a path that is nearly degenerate in free
energy. It intersects the origin at $z=z_2$. The difference between the
two paths is called a ``droplet". It has a length $\tau$ and a width $\Delta 
\sim \tau^\zeta$.}
\end{figure}

\begin{figure}
\caption{ A loop diagram describing the renormalization of the pinning 
strength $u$ given in 
Eq.~(4.25). The three-point function
$G^{(2,1)}$ is given by the convolution of a vertex (the black triangle)
 and three $G^{(1,1)}$'s ( lines with arrows). The mode-coupling approach
Eq.~(A21) approximates the vertex by the ``bare'' vertex, i.e.,  two spatial derivatives (see Appendix A).}
\end{figure}


\begin{references}
\bibitem{note.1} In carrying out this calculation, we used an explicit
Gaussian form for the crossover function $\tg$ (see Appendix A). We
believe the conclusion that there is no singular renormalization of 
$\Delta F_0$ should be insensitive to the specific form of $\tg$.
\bibitem{corr.note} To simplify notation, we define our energy scale such
that the mean free energy is zero, i.e., $\overline{F_0(\bx,t)} = 0$.
\end{references}
\end{document}